\documentclass[boxit]{JAC2003}

\usepackage{graphicx}
\usepackage{booktabs}
\usepackage{verbatim}
\usepackage{subcaption}


 {\usepackage[utf8]{inputenc}}           % switch to utf8

\usepackage[USenglish]{babel}

\begin{document}

\title{DEMIRCI: An RFQ Design Software\thanks{Work supported by TUBITAK with 114F106 project number}}

\author{B. Yasatekin\thanks{byasatekin@ankara.edu.tr}, G. Turemen, Ankara University, Department of Physics \& SANAEM, Ankara, Turkey\\
        E. Celebi, Bogazici University, Department of Physics, İstanbul, Turkey\\
        G. Unel, University of California at Irvine, Department of Physics and Astronomy, Irvine, USA\\
        O. Cakir, Ankara University, Department of Physics, Ankara, Turkey}

\maketitle

\begin{abstract}
   The development and production of radio frequency quadrupoles, which are used for accelerating  low-energy ions to high energies, continues since 1970s. The development of RFQ design software packages, which can provide ease of use with a graphical interface, can visualize the behavior of the ion beam inside the RFQ, and can run on both Unix and Windows platforms, has become inevitable due to increasing interest around the world. In this context, a new RFQ design software package, DEMIRCI, has been under development. To meet the user expectations, a number of new features have been recently added to DEMIRCI. Apart from being usable via both graphical interface and command line, DEMIRCI has been enriched with beam dynamics calculations. This new module gives users the possibility to define and track an input beam and to monitor its behavior along the RFQ. Additionally, the Windows OS has been added to the list of supported platforms. Finally, the addition of more realistic 8 term potential results has been ongoing. This note will summarize the latest developments and results from DEMIRCI RFQ design software.
\end{abstract}

\section{Introduction}
The task of designing  a radio frequency quadrupole (RFQ) necessitates the usage of dedicated design software packages~\cite{TepilKap}. Recently, the DEMIRCI project was started to supplement the existing such softwares~\cite{Lidos, Parmteq}, with the goal of benefiting from modern concepts such as ROOT~\cite{root} environment and OO programming in order to make it easy to use and maintain. The initial version of this new program has been discussed previously ~\cite{demirint}. During last year, the new version of DEMIRCI (v1.9) has been enriched with new design features which constitute the main subject of this note~\cite{get_code}. One of the new features is the ability of running beam dynamics simulations based on user selected beam types for a user selected number of particles. These simulations are expected to make the overall design experience more realistic by monitoring beam behavior in phase and normal spaces along the RFQ. One other feature of the new version is the support of Windows platform as a base operating system alongside Linux and Mac OSX. Amongst additional enhancements that will be also discussed below, one can count the realistic 2D profile definition of the RFQ cell and subsequent usage of SuperFish~\cite{superfish} for detailed studies.

\section{New Developments}
\begin{figure}[!htb]
    \centering
     \includegraphics*[width=\columnwidth]{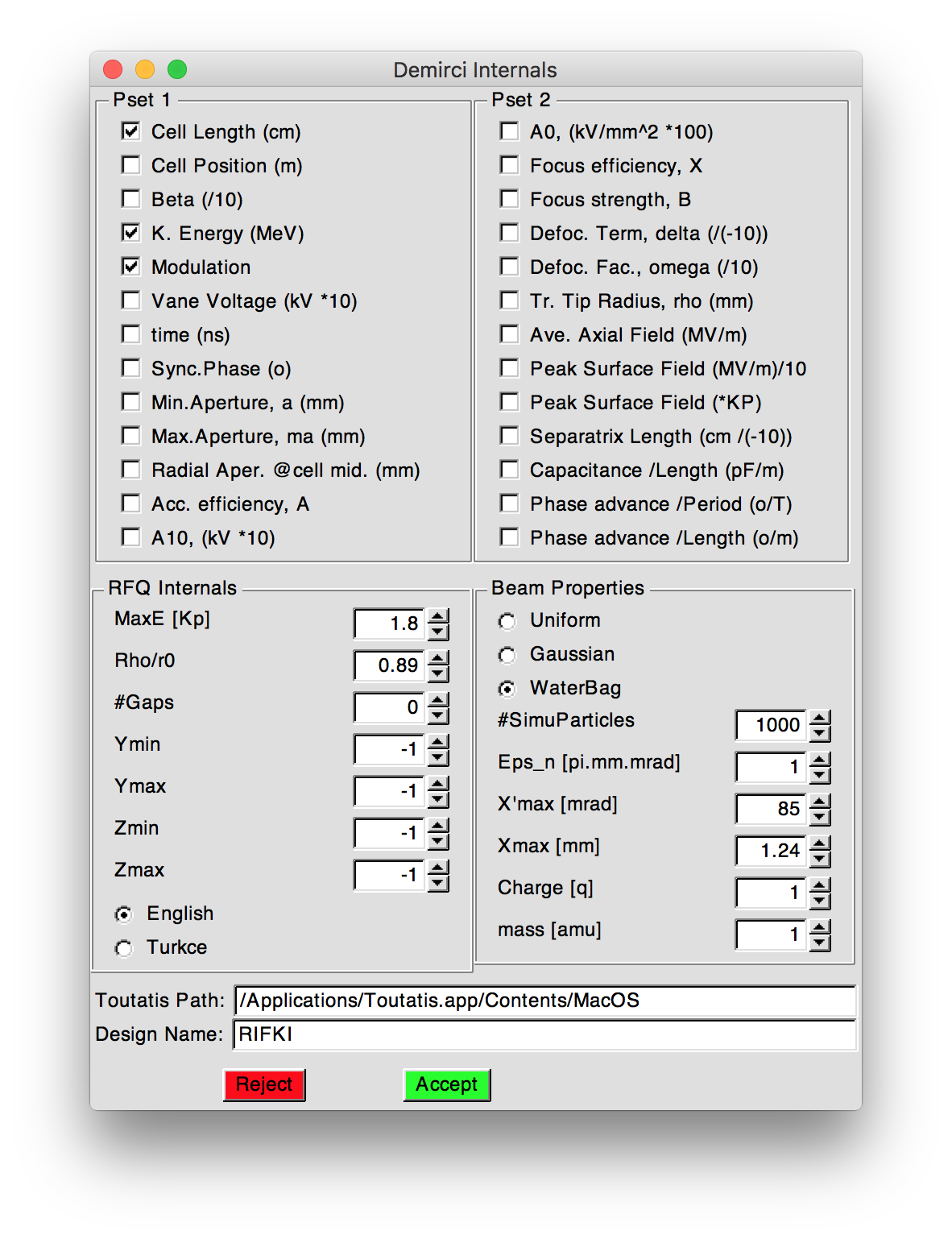}
    \caption{The enhanced setup window with beam parameters and their default values.}
    \label{setupbox}
\end{figure}

The new developments mainly focus around the addition of the beam dynamics calculations and the addition of more terms to the potential function. As the latter is an ongoing effort, this note will focus on the former, the new supported platform and the enhanced interface to SuperFish. First of all, to define the particle beam at the entrance of the accelerator cavity, a new set of internal parameters were added to DEMIRCI. These are the beam distribution type, the beam emittance, the charge and mass of the particles under consideration and the total number of macro-particles to be used in the simulations. The enlarged setup window to set these new variables can be seen in Fig.~\ref{setupbox} with their default values. One can also notice that language localization has been added to DEMIRCI, Turkish being the second language supported after English.

\subsection{Beam Dynamics}
\begin{figure}[!htb]
    \centering
     \includegraphics*[width=80mm]{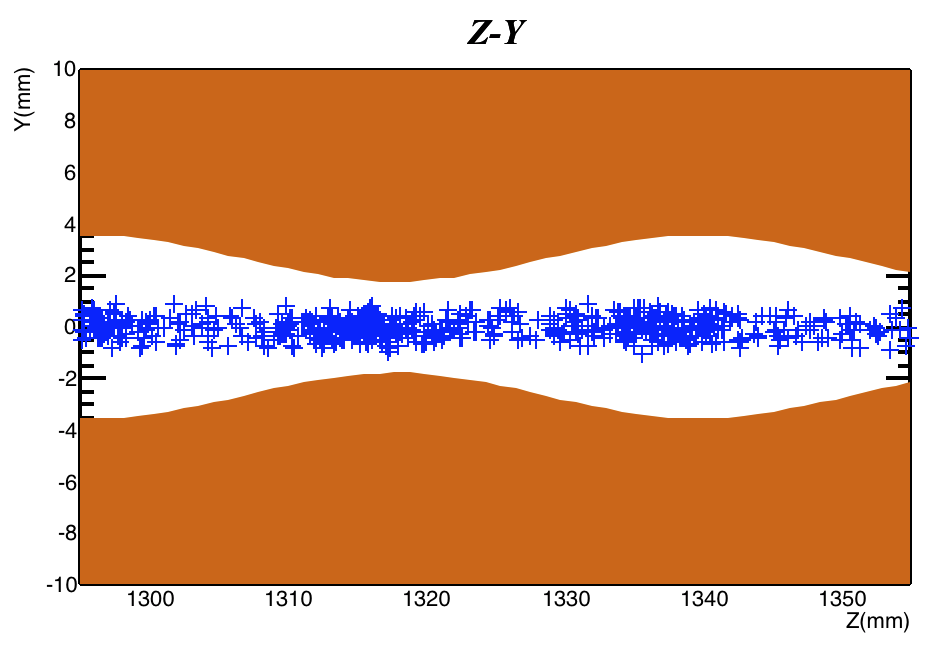}
      \includegraphics*[width=80mm]{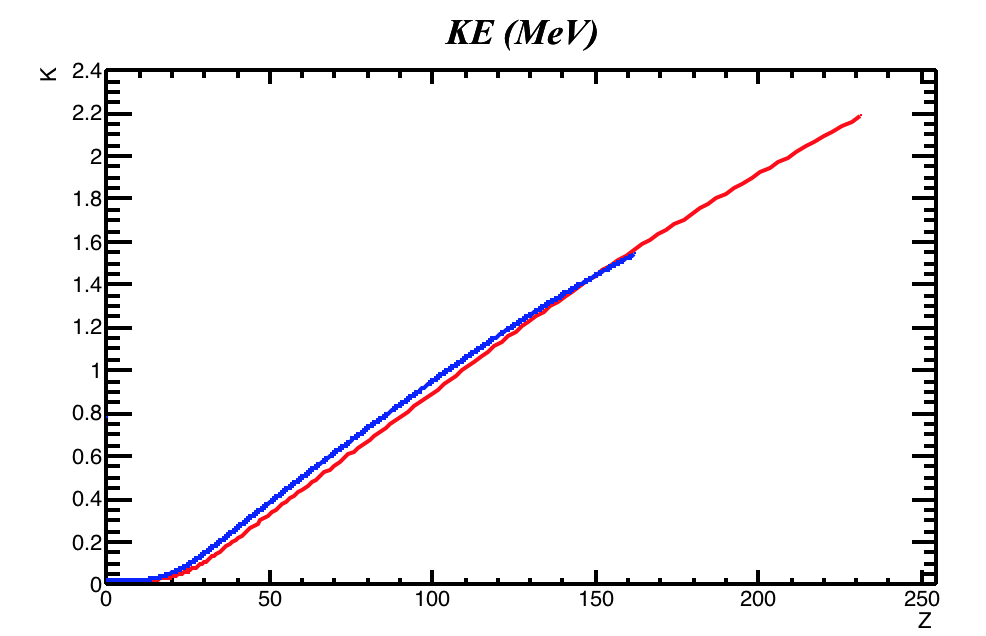}
    \caption{Top: bunching of particles in an RFQ becomes visible. Bottom: Average kinetic energy (in blue) of the macro particles obtained from dynamic equations in blue compared to estimations (in red) from averaged design equations in red. }
    \label{beam2}
\end{figure}

At the initialization step,  a number of macro particles are randomly generated in the phase space according to beam definitions just at the entrance of the RFQ. For individual particle motion, between the time based and position based approaches, DEMIRCI employs the former as it is more convenient for the user. In this approach, a small enough time step ($\delta t$) is selected to calculate the motion of particles according to the electric field ($\vec{E}$) values at the particle position, velocity and time. The relevant equations for position, velocity and kinetic energy at the next time interval ($t+\delta t$) are~\cite{toutatis}:

\begin{eqnarray}
\vec{x}^{t+\delta t}&=&\vec{x}^t + \vec{\beta}^t c \delta t + {\frac{1}{2}} {\frac{Q}{\gamma^t M}} \vec{E}^t \delta t^2  \quad , \\
{\gamma}^{t+\delta t} \vec{\beta}^{t+\delta t} &=& \gamma^t  \vec{\beta}^t  + {\frac{1}{2}} {\frac{Q}{cM}} \delta t (\vec{E}^t +\vec{E}^ {t+\delta t}) \quad , \\
E_k^{t+\delta t} &=& (\gamma^{t+\delta t}-1)Mc^2 \quad ,
\end{eqnarray}
where quantities with superscript $t$ represent their values at time $t$, whereas $Q$ and $M$ are the charge and mass of the macro-particle. The DEMIRCI library has been enlarged with functions that calculate and plot the particle trajectories according to the above equations.
\begin{figure}[!htb]
    \centering
     \includegraphics*[width=\columnwidth]{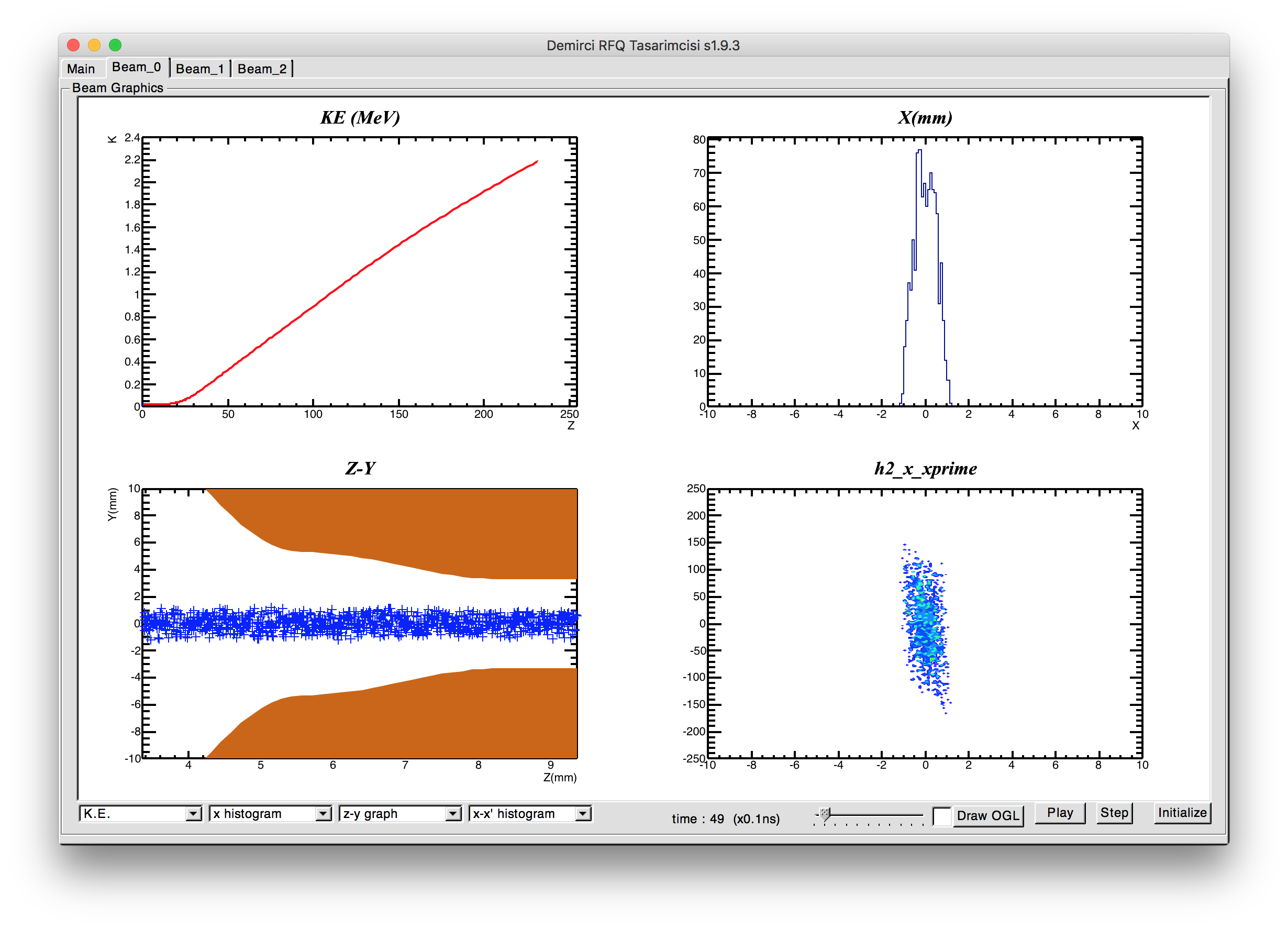}
    \caption{The new beam dynamics tab of DEMIRCI.}
    \label{beamplots}
\end{figure}
For displaying the information, four beam dynamics tabs were added to the main DEMIRCI window. These become accessible once an RFQ description is loaded and its overall behavior is calculated using previously described methodology ~\cite{demirpaper}. As seen in Fig.~\ref{beamplots}, each tab can only show four plots at a time, however up to four tabs were made available to the user for monitoring multiple distributions. The selection is made using the selection drop-down lists at the lower left side of the beam window. The bottom right hand side of the same window contains the buttons for initializing, starting and pausing the simulations which can also be executed stepwise. The selected time step of 0.1 ns and the large number of particles used in beam dynamics could be more than a CPU can handle. To overcome this problem, the graphics refresh rate is made user selectable using the scroll bar at the bottom center of the window. By adjusting it the user can play the simulation rapidly (up to 10 times) and slow it down when needed to examine in detail the important stages. One final visualization aid is the so called the OGL button, next to the speed adjustment slider. It was noticed that although the phase space plots like X-X' are great indicators while addressing issues related to beam dynamics plots, it is hard to visualize the overall beam behavior in 3D. Therefore the OGL button displays a new window in which 3D particles (with exaggerated volumes) are displayed for the relevant simulation step.
Currently work focuses on using this tool to evaluate the performance of various RFQ designs and comparing the results of multi particle simulations to design equation results as can be observed in Fig.~\ref{beam2}.

\subsection{2D Cross Section}
\begin{figure}[!htb]
    \centering
     \includegraphics*[width=\columnwidth]{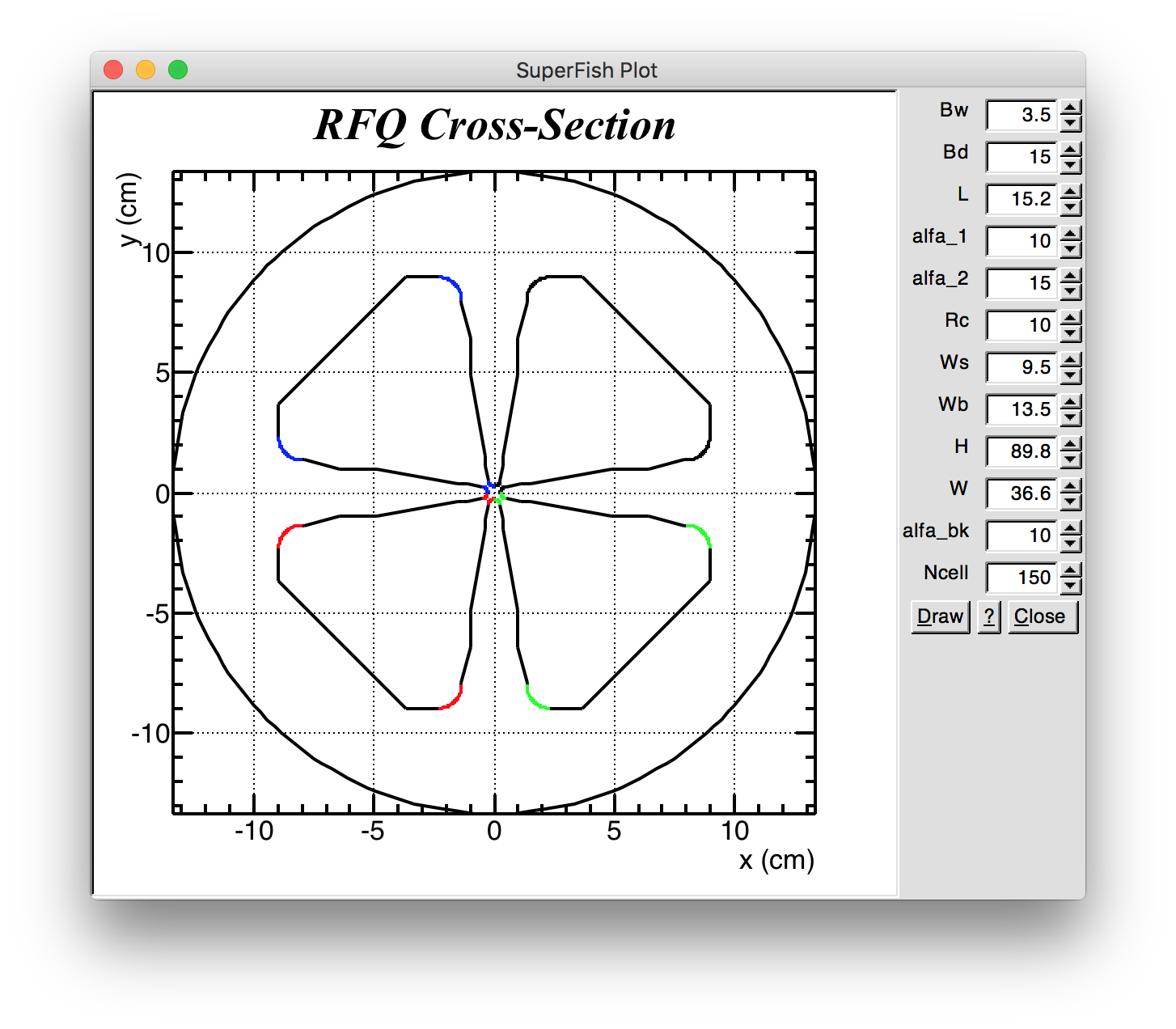}
    \caption{The RFQ cross section window.}
    \label{sfish}
\end{figure}

In order to use the existing 2D EM design software SuperFish, the relevant parts of DEMIRCI have been updated as seen in Fig.~\ref{sfish}. It is now possible to define the x-y cross section of any cell in a realistic way by specifying a total of 12 new parameters. These are vane skirt angles and distances together with connection angles to the RFQ wall. Their exact definitions can be found in the help menu accessible via the "?" button. The drawing routine also produces an output file that can directly be read by SuperFish to calculate quantities like the power loss on the walls.

\subsection{A New  Platform}
\begin{figure}[!htb]
    \centering
     \includegraphics*[width=\columnwidth]{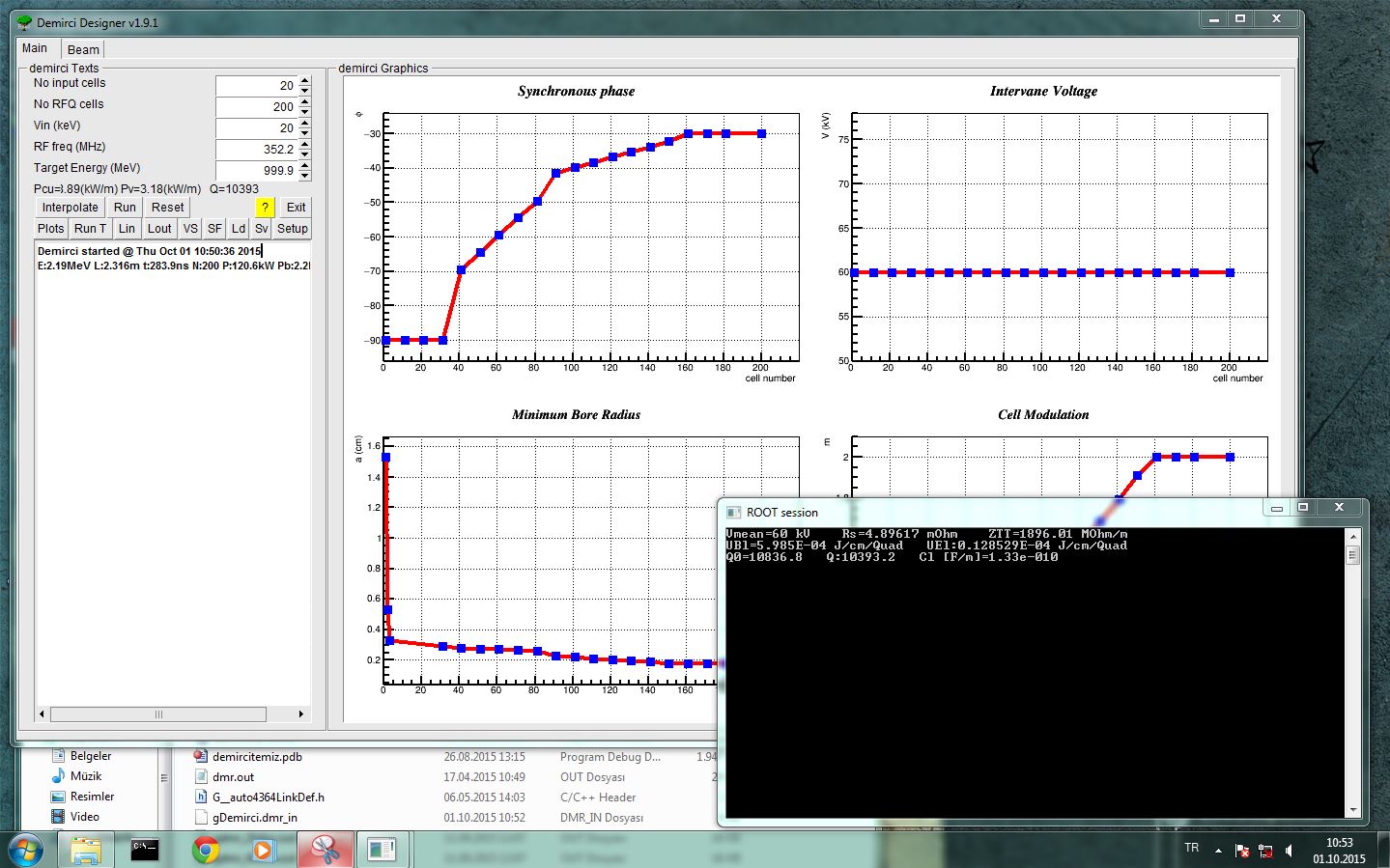}
    \caption{A screenshot of DEMIRCI v1.9, on Windows OS.}
    \label{demwin}
\end{figure}

Previously, DEMIRCI's base operating systems were only various Unix based platforms, such as  Linux distributions and Mac OSX. A recent web based poll amongst the members of the accelerator community revealed a strong request to support the Windows operating system. To match this request, DEMIRCI has been ported to Windows with both graphical and command line interfaces. The minimum prerequisite libraries to run the Windows version of DEMIRCI are:  Microsoft Visual Studio C++ version 2010 Express~\cite{vs} and ROOT version 5.34/26 ~\cite{root}. These software packages can be obtained free of charge. A screenshot of DEMIRCI running on Windows can be seen in Fig.~\ref{demwin}. With this additional OS, it has become possible to interoperate DEMIRCI with other accelerator related software packages which are available on only Windows platforms, such as LIDOS ~\cite{Lidos} and SuperFish ~\cite{superfish}.

\section{Conclusion}
The development of the ROOT based  RFQ design software DEMIRCI continues with the addition of many novelties like multi particle beam dynamics, new operating system and interoperability with other similar software. Performance comparison with similar software packages and the finalization of the 8-term potential implementation are also ongoing with the ultimate goal of delivering a reliable, user friendly and multi-platform solution to the RFQ design problem within 2016.

\section{acknowledgment}
The authors would like to thank TUBITAK for their support under project number 114F106 and Dr. Sezen Sekmen for a careful reading of the manuscript.

\end{document}